\documentclass[preprint,times]{elsarticle}
\pdfoutput=1
\usepackage{amsmath}  
\usepackage{amssymb}
\usepackage{subfigure}
\usepackage{ecrc}
\usepackage{rotating}
\usepackage{xcolor}
\usepackage{latexsym}
\usepackage{bm}
\usepackage{floatrow}
\newfloatcommand{capbtabbox}{table}[][0.45\textwidth]

\volume{}

\jid{}
\firstpage{1}
\journalname{}
\jnltitlelogo{}
\runauth{}

\newcommand{\be}{\begin{equation}}
\newcommand{\ee}{\end{equation}}
\newcommand{\ba}{\begin{eqnarray}}
\newcommand{\ea}{\end{eqnarray}}

\newcommand{\al}{\alpha}
\newcommand{\bt}{\beta}
\newcommand{\gm}{\gamma}
\newcommand{\Dl}{\Delta}
\newcommand{\kp}{\kappa}

\begin{document}
\begin{frontmatter}

\title{Intergranular stress distributions in polycrystalline aggregates\\ of irradiated stainless steel}

\author[CEA]{J.~Hure\corref{cor1}}
\author[JSI]{S.~El~Shawish}
\author[JSI]{L.~Cizelj}
\author[CEA]{B.~Tanguy}

\cortext[cor1]{Corresponding author}
\address[CEA]{CEA, Universit\'{e} Paris-Saclay, DEN, Service d'\'Etudes des Mat\'{e}riaux Irradi\'{e}s, 91191 Gif-sur-Yvette cedex, France}
\address[JSI]{Jo$\check{z}$ef Stefan Institute, SI-1000, Ljubljana, Slovenia}

\begin{abstract}
In order to predict InterGranular Stress Corrosion Cracking (IGSCC) of
post-irradiated austenitic stainless steel in Light Water Reactor (LWR)
environment,
reliable predictions of intergranular stresses are required.
Finite elements simulations have been performed on realistic
polycrystalline aggregate with a recently proposed physically-based crystal plasticity constitutive equations validated for neutron-irradiated austenitic
stainless steel.  Intergranular normal stress probability density
functions are found with respect to plastic strain and irradiation level,
for uniaxial loading conditions. In addition, plastic slip activity jumps at grain boundaries are also presented.  Intergranular normal stress distributions describe, from a
statistical point of view, the potential increase of intergranular
stress with respect to the macroscopic stress due to grain-grain
interactions. The distributions are shown to be well described by a master
curve once rescaled by the macroscopic stress, in the range of
irradiation level and strain considered in this study. The upper tail
of this master curve is shown to be insensitive to free surface effect, which is relevant for IGSCC predictions, and also relatively insensitive to small perturbations in crystallographic texture, but sensitive to grain shapes.
\end{abstract}


\end{frontmatter}

\section{Introduction}
\label{introduction}

Internals structures of Light Water Reactors (LWR) are made mainly of
austenitic stainless steels, known for their good mechanical
properties such as ductility and toughness, and their resistance to
corrosion. Under neutron irradiation, a degradation of these
properties is observed, such as a drastic decrease of fracture
toughness and a susceptibility to stress corrosion cracking (SCC)
\cite{chopra1,chopra2,fukuya2013}. This last phenomenon is referring
to as Irradiation Assisted Stress Corrosion Cracking (IASCC), as it
affects materials under irradiation initially non-sensitive to SCC. IASCC has
been observed since the 1980's in both Boiling Water Reactors (BWR)
and Pressurized Water Reactors (PWR), with the appearance of
intergranular cracks for example in core shrouds in BWRs and
baffle-to-former bolts in PWRs \cite{IASCC_IAEA,fieldIASCC}. Several
researches have been conducted since then to assess the key parameters
affecting both initiation and propagation of intergranular cracks in
LWR's environment, parameters than can be divided into three groups:
stress level \cite{chopra2,freyer,conermann,takakura2007,nishioka2008}, mechanical behaviour of the material and
its chemical composition (which includes initial microstructure and
chemical composition, and irradiated defects and segregation
\cite{stephenson2014}), and LWR's environment
\cite{katsura1993,fujii2010}.

In recent years, different models have been proposed to predict IASCC of austenitic
stainless steels, or more generally InterGranular Stress Corrosion Cracking (IGSCC). In some studies, slip
localization, \textit{i.e.} strongly heterogeneous deformation field at the grain scale has been argued to be responsible for IGSCC crack
initiation, leading to models based on dislocation pile-up theory or
refined slip bands modelling \cite{sauzay2013}. Other recent studies
of SCC cracking are based on simulations of polycrystalline aggregates
with cohesive zone models \cite{simono2014}, leading to both
prediction of initiation and propagation. These models using standard crystal plasticity constitutive equations assume implicitly that rather homogeneous deformation field at the grain scale can be sufficient to assess intergranular cracking. Such models have been used
for example to predict SCC of Zircaloy in iodine environment
\cite{musienko2009} or SCC of cold-worked austenitic steels
\cite{couvant2013}. These fully coupled models require both an
accurate crystal plasticity constitutive law and grain boundaries
modelling in order to provide quantitative predictions. As crystal
plasticity models are only recently available in the literature for irradiated materials
\cite{de2014,xiao2015}, and especially irradiated stainless steels
\cite{barton2013,xuhanconf,phdHan}, such simulations have not yet been performed
so far for irradiated stainless steel.

These fully coupled models show some drawbacks such as high
computational cost due to crack propagation and rather inaccurate intergranular stress prediction due to cohesive zone modeling \cite{shawish}. Therefore, an intermediate approach is considered here
following the work of Diard \textit{et al.}
\cite{diard2002,diard2005}, based on the computation of accurate
intergranular (normal) stress distributions for irradiated stainless
steels for uncracked polycrystalline aggregate, as normal stress at grain boundaries is assumed to be the key parameter for intergranular cracking.  Contrary to other studies \cite{diard2002,gonzalez2014} focusing on relations between
intergranular stress and orientation of the boundary with respect to
the loading direction or to the mismatch of deformations between
adjacent grains, full distributions for statistically large number of
grains with different orientations are assessed. Combined with an empirical or experimental criterion for grain boundary strength that may depend on oxidation time and irradiation level, such uncoupled modelling could in principle be an efficient tool to predict critical macroscopic stress above which intergranular cracks are expected to be initiated at the grain scale, or intergranular surface cracking density. This approach is particularly relevant in the case of SCC of post-irradiated or heavily cold-worked austenitic stainless steel \cite{raquet} in PWR nominal environment (compared to BWR oxygenated environment), where low propagation rates enables rather weak interactions between intergranular cracks, and thus the appearance of numerous surface cracks, as shown for example in \cite{stephenson2014} and \cite{lemillier}, for neutron-irradiated and proton-irradiated stainless steel, respectively.

In the first part of the paper, recently proposed crystal plasticity
physically-based constitutive equations for neutron-irradiated austenitic stainless steel \cite{xuhanconf,phdHan} that have been used in this study are described. 
In a second part, polycrystalline simulations based on a realistic aggregate are presented, leading to the stress
distributions at grain boundaries as a function of strain and
irradiation level. The effects of free surface, deviations from random crystallographic texture and equiaxed grain shapes are assessed. As a conclusion, and based on these numerical results, a methodology is proposed to obtain a statistical modelling of IGSCC initiation. The range of validity of such modelling is finally discussed.

\section{Numerical modelling}
\subsection{Crystal plasticity constitutive model}
\label{sec:cpmodel}

A degradation of mechanical properties of irradiated austenitic
stainless steels
is commonly ascribed to the formation of high density nano-sized
irradiation defect clusters, mainly interstitial Frank loops
\cite{frank}, preventing the motion of dislocations thus leading to
hardening, and also reducing strain-hardening capabilities.
To account for those effects, a crystal plasticity constitutive model developed in \cite{xuhanconf,phdHan} is described, with the dislocation and Frank loop density-based
evolution laws.

To describe the plastic behaviour of single crystalline material, in
general two laws are considered: a shear flow law that activates a
slip system and determines its slip rate, and a hardening law that
describes the change of slip activation with applied slip by taking
into account the evolution of various defects created within the
material.
The shear flow adopted here for non-irradiated and irradiated conditions
is of visco-plastic type and represents isotropic hardening,
\be
  \dot{\gm}^\al=\left\langle\frac{|\tau^\al| - \tau_c^\al}{K_0}\right\rangle^n {\rm sign}(\tau^\al),
  \quad\hbox{with}\quad\langle x\rangle=\left\{ \begin{array}{ll}
	x &; x>0\\
        0 &; x\le 0
	\end{array}\right.
  \label{eq_gm}
\ee
where $\gm^\al$ is shear strain in slip system $\al$ ($\al=1\ldots 12$
for Face-Centered-Cubic lattice) and $\tau^\al$ and $\tau_c^\al$ are
respectively the resolved shear stress and critical resolved shear
stress. Parameters $K_0$ and $n$ regulate the viscosity of the shear
flow.

The critical resolved shear stress is additively decomposed into
components that contribute to the hardening \cite{mecking},
\be
  \tau_c^\al = \left\{ \begin{array}{ll}
  \tau_0+\mu\sqrt{\sum\limits_{\bt=1}^{12} a^{\al\bt}r_D^\bt} &; \hbox{non-irradiated} \\
  \tau_0+ \tau_a\exp\left(-\frac{|\gm^\al|}{\gm_0}\right) +
  \mu\sqrt{\sum\limits_{\bt=1}^{12} a^{\al\bt}r_D^\bt}+\mu\al_L\sqrt{\sum\limits_{p=1}^4 r_L^p} &; \hbox{irradiated}
  \end{array}\right. 
  \label{eq_tc}
\ee
The hardening law for non-irradiated steels accounts only for
dislocation density evolution. Here, $r_D^\al$ is normalized
dislocation density in slip system $\al$ (normalization factor $b^2_D$, with Burgers vector $b_D = 2.54.10^{-10}$m), $\tau_0$ is lattice friction
stress that remains constant for a given temperature, while $\mu$ and
$a^{\al\bt}$ are respectively the macroscopic shear modulus and
$12\times 12$ matrix (with 6 independent parameters) of long-range
interactions between dislocations.
In the irradiated material, additional hardening is expected due to the 
presence of Frank loops. Here, $r_L^p$ is a normalized Frank loop
density in slip plane $p$ (normalization factor $b^2_L \phi_L$, with Burgers vector $b_L = 2.08.10^{-10}$m and $\phi_L$ the mean size of Frank loops that depends on irradiation level, see \cite{xuhanconf}) and $\al_L$ sets the relative contribution of Frank loops to hardening.
To account for a dislocation unlock mechanism \cite{phdHan,xuhanconf}, the
additional phenomenological term has been proposed along with a dose
dependant shear stress $\tau_a$ required to unlock the dislocations
and coefficient $\gm_0$ to adjust the speed of the avalanche after
unlocking the dislocations.

The evolution of dislocation density is modeled with a multiplication
and an annihilation term \cite{mecking},
\be
  \dot{r}_D^\al = \left\{ \begin{array}{ll}
  \left(\frac{1}{\kp}\sqrt{\sum\limits_{\bt=1}^{12} b^{\al\bt}r_D^\bt}-G_c r_D^\al\right)|\dot{\gm}^\al| &; \hbox{non-irradiated}\\
  \left(\frac{1}{\kp}\sqrt{\sum\limits_{\bt=1}^{12} b^{\al\bt}r_D^\bt}+
  \frac{1}{\kp}\sqrt{K_{dl}\sum\limits_{p=1}^4 r_L^p}-G_c r_D^\al\right)|\dot{\gm}^\al| &; \hbox{irradiated}
  \end{array}\right. 
  \label{eq_rd}
\ee
where $b^{\al\bt}$ is a matrix of interactions between dislocations,
being of the same shape as $a^{\al\bt}$. Parameter $\kp$ is
proportional to the number of obstacles crossed by a dislocation
before being immobilized and $G_c$ is a proportional factor that
depends on the annihilation mechanism of dislocation dipoles. The
irradiation effects are modeled by adding a term to the multiplication
part, with $K_{dl}$ being a coefficient of effective interaction
between dislocations and Frank loops.

The evolution of Frank loop density in irradiated steels is modeled by Eq.~\ref{eq_rl} proposed in 
\cite{krishna10}
\be
  \dot{r}_L^p = -A_L (r_L^p-r_L^{sat})\left(\sum_{\al\in {\rm plane}\ p}^3\!\!\!\! r_D^\al\right) 
  \left(\sum_{\al\in {\rm plane}\ p}^3\!\!\!\! |\dot{\gm}^\al|\right)
  \label{eq_rl}
\ee
where $A_L$ is the annihilation dimensionless area (rescaling factor $b_L^3/\phi_L$) of Frank loops and $r_L^{sat}$ is
a stabilized value of normalized defect density which depends on the
irradiation dose. Since scanning of Frank loops by mobile dislocations
occurs only within the plane of the loop, only slipping in this plane
can contribute to the evolution of defect density ($\al\in {\rm
plane}\ p$).

Anisotropic elasticity is finally considered, with non-zero parameters
of the elastic fourth order tensor $C_{11}=C_{22}=C_{33}$,
$C_{12}=C_{13}=C_{23}$ and $C_{44}=C_{55}=C_{66}$ in Voigt notations.

\begin{table}[!h]
\begin{center}
\begin{tabular}{c|c}
\hline
\hline
$C_{11}$ & 199\,GPa  \\
$C_{12}$ & 136\,GPa  \\
$C_{44}$ & 105\,GPa  \\
\hline
$K_0$ & 10\,MPa  \\
$n$ & 15  \\
\hline
$\tau_0$ & 88\,MPa  \\
\hline
$\mu$ & 65615\,MPa  \\
$G_c$ & 10.4  \\
$\kappa$ & 42.8  \\
\hline
$a_1$ & 0.124 \\
$a_2$ & 0.124 \\
$a_3$ & 0.07 \\
$a_4$ & 0.625 \\
$a_5$ & 0.137 \\
$a_6$ & 0.122 \\
\hline
$b_i$ &  $1 - \delta_{i1}$     \\
\hline
\hline
\end{tabular}
\begin{tabular}{c|c|c|c|c|c|c}
\hline
\hline
    & 0\,dpa & 0.8\,dpa & 2\,dpa & 3.4\,dpa & 13\,dpa & \\ 
\hline
\hline
$r_D^0$ & $5.38\,10^{-11}$ &  $4.54\,10^{-11}$ &  $3.66\,10^{-11}$ &  $2.97\,10^{-11}$ &  $1.03 \,10^{-11}$ & \\
$r_L^0$ & - &  $2.29\,10^{-6}$ &  $4.72\,10^{-6}$ &  $5.04\,10^{-6}$ &  $4.9 \,10^{-6}$ & \\
\hline
$\alpha_{L}$ & - & 0.21 & 0.44 & 0.49 & 0.57 & \\
$K_{dl}$ & - &  \multicolumn{4}{c|}{$0.25\,10^{-6}$} & \\
$A_L$ & - & -  & $4.48\,10^{8}$ & $5.62\,10^{8}$ & $5.55\,10^{8}$ & \\
$r_L^{sat}$ & - &  $2.29\,10^{-6}$ &  $3.78\,10^{-6}$ &  $3.98\,10^{-6}$ &  $3.23 \,10^{-6}$ & \\
\hline
$\tau_{a}$ & - & - & 50.0\,MPa & 61.3\,MPa & 61.2\,MPa & \\
$\gamma_{0}$ & - &  \multicolumn{4}{c|}{$5\,10^{-3}$} & \\
\hline
\hline
\end{tabular}
\end{center}
\caption{
Parameters of the constitutive law for neutron-irradiated austenitic
stainless steel SA304 at 330$\mathrm{^{\circ}C}$ \cite{xuhanconf}.}
\label{tab:par}
\end{table}

Parameters of the constitutive law have been determined by fitting a
polycrystalline aggregate model response \cite{phdHan,xuhanconf} to
tensile measurements on 304 stainless steel at 330$^{\circ}$C
\cite{pokor04}. This crystal plasticity model is able to reproduce
tensile tests up to 13 dpa, as well as the evolution of dislocation
and Frank loops densities as a function of strain and irradiation
level. The parameters are summarized in Table \ref{tab:par}.

\subsection{Numerical implementations}

The above constitutive law has been implemented into two numerical
codes: Abaqus \cite{abaqus} and Cast3M \cite{castem} with different methods to assess robustness of numerical integration. 

In both implementations finite strain is accounted for by
the usual multiplicative decomposition of the deformation gradient,
\be
  F=F^e F^p.
\ee
The crystal deforms solely through plastic shearing on
crystallographic slip systems from the reference configuration to an
intermediate configuration by the plastic deformation gradient
$F^p$. The elastic deformation gradient, $F^e$, then stretches and
rotates the lattice to bring the crystal to the final configuration.

The evolution of $F^p$ is given by the flow rule,
\be
  \dot{F}^p=L^p F^p
\ee
where $L^p$ is the local velocity gradient assumed to arise only from
the plastic shearing of individual slip systems of the crystal defined
as
\be
  L^p=\sum_{\al=1}^{12}\dot{\gm}^\al N^\al {\quad \rm with \quad}
  N^\al=s^\al\otimes m^\al
\ee
and $s_i^\al$ being a vector lying along the slip direction and
$m_i^\al$ a vector normal to the slip plane of system $\al$.

The elastic strain measure is given by Green-Lagrange strain tensor,
\be
  E=\frac{1}{2}\left((F^e)^T F^e - I\right),
\ee
and the stress measure by Mandel stress tensor,
\be
  M=\det(F^e) (F^e)^T \sigma (F^e)^{-T},
\ee
where $\sigma$ is the true (Cauchy) stress. In this way, the resolved
shear stress, $\tau^\al$, can be calculated as
\be
  \tau^\al=M:N^\al=M:(s^\al\otimes m^\al).
\ee

The two different methods used for numerical integration are briefly sketched in Appendix, focusing
primarily on the integration of the internal variables. The
differences between the two implementations are highlighted. Both implementations (Abaqus and Cast3M) have been shown to give
equivalent results for different crystal orientations. In the
following, results coming from these two different implementations
will not be differentiated.

\subsection{Polycrystalline aggregates}

\begin{figure}[H]
\centering
\subfigure[]{\includegraphics[height = 4cm]{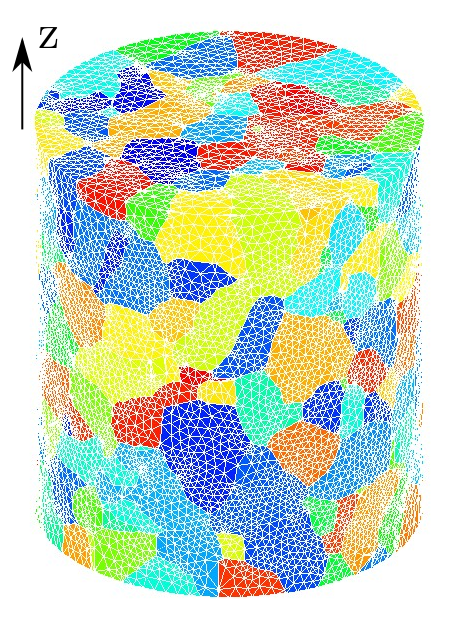}}
\subfigure[]{\includegraphics[height = 4cm]{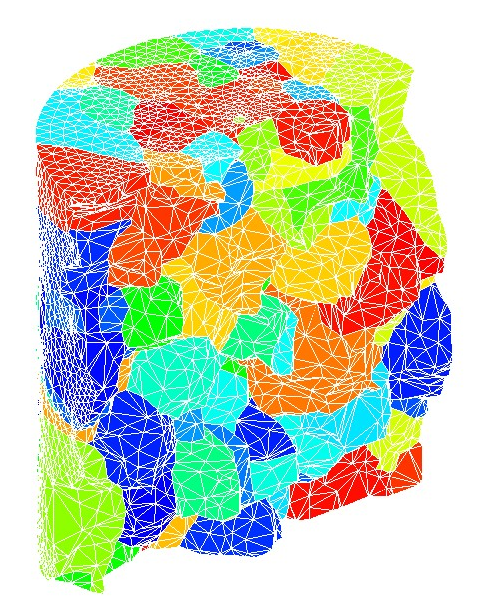}}
\subfigure[]{\includegraphics[height = 4cm]{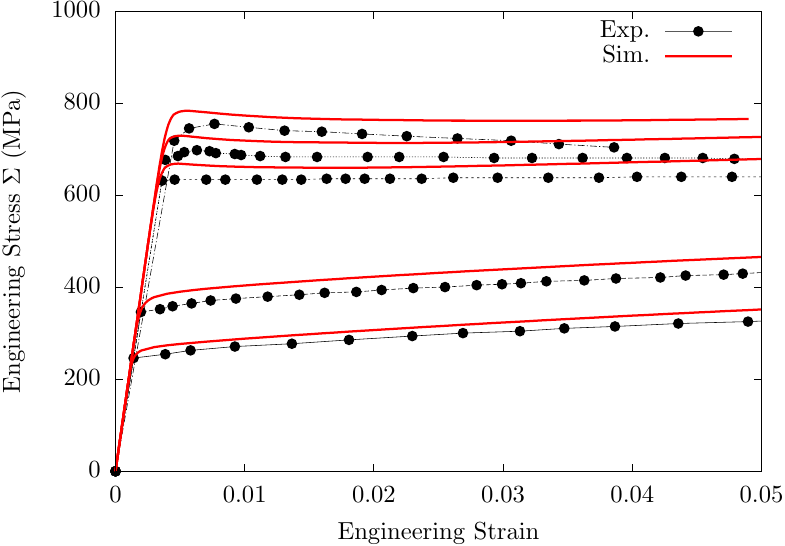}}
\subfigure[]{\includegraphics[height = 4cm]{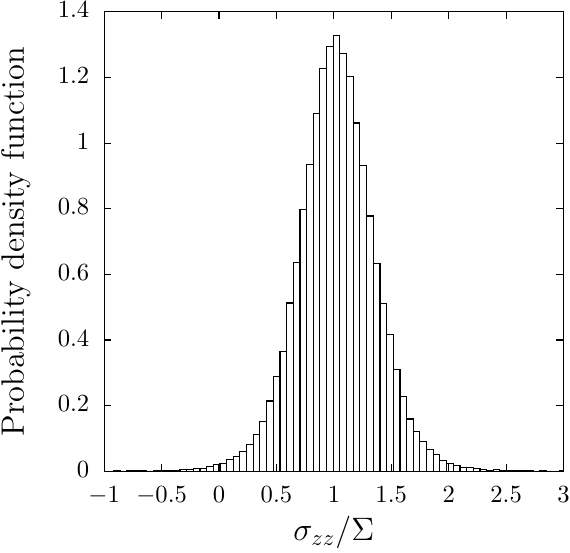}}
\caption{
(a) The mesh used throughout this study, composed of 142k tetrahedral quadratic
elements, has been obtained from X-ray Diffraction Contrast
Tomography (DCT) on an austenitic stainless steel wire of 400$\mu$m
diameter \cite{simono2011}.  Each color correspond to a different
grain, characterised by its crystallographic orientation also given by
DCT. The wire is composed of 377 grains. (b) Details of the interior
mesh of the wire. (c) Tensile Tests: Comparisons between experimental data (Exp.) and
numerical simulations (Sim.) up to 5\% total strain, for different levels of irradiation [0,0.8,2,3.4,13] dpa. (d) Typical distribution of axial stress in the aggregate.}
\label{fig:wiremesh1}
\end{figure}

Polycrystalline aggregate models are generated upon analytic and
realistic grain structures. They are, respectively, built from Voronoi
tessellations with random initial grain orientations and from diffraction contrast tomography
data \cite{king08} of a stainless steel wire specimen (see, {\it
e.g.}, Fig.~\ref{fig:wiremesh1}a). The framework for building a finite
element model of grains in realistic spatial structures is described
in detail in \cite{simono2011}. In both cases, the grains have been
assigned the material model introduced in Section
\ref{sec:cpmodel} while the grain boundaries have not been modeled
explicitly. 

\begin{figure}[H]
\begin{floatrow}
\capbtabbox{%
  \begin{tabular}{l l l l l}
& & & & \\ 
\hline
\hline
model     & type & grains   & elements & orientations\\
\hline
\hline
VORO -- 216   & analytic &   216    &  38350 & random   \\
VORO -- 343   & analytic &   343    &  64331 & random   \\
Wire coarse  & realistic &   377    &  141968 & realistic  \\
Wire fine  & realistic &   377    &  796105 & realistic  \\
\hline
\hline
\end{tabular}
}{%
  \caption{Finite element models of polycrystalline aggregates with 
    the corresponding numbers of grains and elements. Initial 
    crystallographic grain orientations are also shown. Note that 
    random orientations provide no texture.}%
}
\ffigbox{%

\includegraphics[height = 3cm]{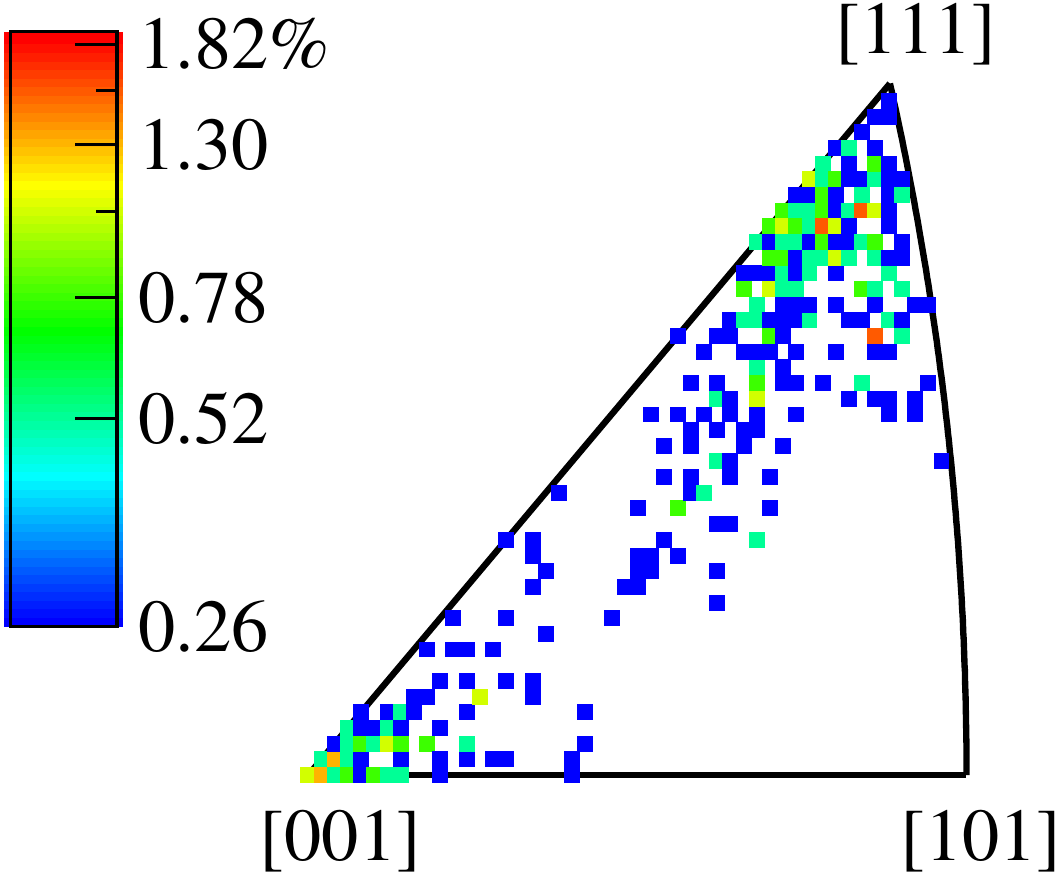}
}{%
  \caption{Inverse pole figure for rolling (tensile) model direction showing the
initial distribution of the crystallographic orientations in the wire
model. Colors denote a percentage of the grains in the model sharing
the same pole value.}%
}
\end{floatrow}
\end{figure}

In all considered models the grains are meshed by quadratic
tetrahedral elements. Two mesh densities are tested for each
Voronoi and wire topology to estimate finite element size effects. In this
respect, meshes with up to 64k and 800k elements have been used in the
analyses for Voronoi and wire geometries, respectively (Tab.~2). Wire model shows crystallographic texture, as shown on Fig.~2.

Regarding boundary conditions, an incremental tensile displacement is
applied along the axial ($Z$) axis to all the nodes on the top
surface, while the nodes on the bottom surface are constrained to have
zero axial displacement. Nodes on the lateral surface are not
constrained so as to study the free surface effect. The applied
nominal strain is $\Delta z/z_0=0.05$ and the strain rate is in the
range [$10^{-4}-10^{-3}$ s$^{-1}$]\footnote{Parameters $K_0$ and $n$ of
the shear flow rule have been chosen such as to reproduce the weak
viscous behaviour of austenitic stainless steel at
330$\mathrm{^{\circ}C}$ \cite{xuhanconf}, thus there is no effect of
strain rate on stresses in the range used in this study.}.

Figure \ref{fig:wiremesh1}(c) compares the calculated and measured
engineering tensile curves on 304L stainless steel \cite{pokor04} at
five levels of irradiation. Simulations are performed up to 5\% plastic strain, as IGSCC initiation of irradiated stainless steel is usually observed close or below the yield stress. The curves show typical dose dependence
for austenitic stainless steel \cite{clear}: with increasing
irradiation the yield stress is also increased and a pronounced
softening is observed just after the yield point at doses equal or
higher than 2.0 dpa. The corresponding calculations have been
performed on a wire model up to 0.05 strain using the material
parameters from Table \ref{tab:par}. A qualitatively good agreement
with the experiment is observed for all doses. As macroscopic convergence with respect to the number of grains and mesh density is achieved for the wire model (see section~4), the slightly stiffer
response at higher levels of irradiation can be attributed to the
initial (as measured) texture of the wire which is a consequence of
the manufacturing process. Distribution of axial stress in the aggregate is given in Fig.~1d, where a typical gaussian shape is obtained.

\section{Intergranular normal stress distributions}
\label{interdis}
Simulations performed on a wire austenitic stainless steel
aggregate with physically-based crystal plasticity constitutive
equations allow to obtain probability density functions (thereafter
noted \textit{pdf}) of normal stress $\sigma_{nn}$ between adjacent
grains as a function of strain and irradiation. As brittle cracking of
grain boundaries is related to the level of normal stress, these pdf
are believed to be a key ingredient towards IGSCC modelling. Results
presented in this section correspond to the \textit{coarse wire}
finite element model. Parameters that may have an influence on
intergranular normal stress distribution - such as free surface,
crystallographic texture, grain shapes - are assessed in the next
section.

For each pair of tetrahedral elements defining a boundary between two
grains, the stress tensors $\bm{\underline{\sigma}}$ at the closest
Gauss points near the boundary are obtained, then converted to normal stresses
knowing the normal $\underline{n}$ of the grain boundary facet $\sigma_{nn}= \underline{n}. \bm{\underline{\sigma}} .  \underline{n}
$, and finally averaged to yield a single value. It should be noted that the value of the intergranular normal stress strongly depends on the cosine of the angle between the normal to the grain boundary and the loading direction, which is almost uniformly distributed for the polycrystalline aggregates used in this study\footnote{See inset Fig.~9b and Section 4.2 for discussion.}, and is bounded by the macroscopic uniaxial stress in the absence of mismatch effects between adjacent grains. As
elements of different sizes are used in the mesh, the occurrence of
$\sigma_{nn}$ is weighted by the surface of the grain boundary facet on which it
was obtained for the computation of the \textit{pdf}. One may note
however that such rescaling has only a small effect on the results.

The methodology used here to obtain intergranular normal stress
differs from those used in \cite{diard2002} and
\cite{gonzalez2014}. In the former, additional Gauss points are
considered in elements close to grain boundaries, while cohesive zone
model elements are added to the model in the latter to obtain stresses
precisely at grain boundaries. The rather simple method used in this
study is however shown to be accurate enough to get the global shapes of the distributions in
Section~\ref{sensitivitystudy} upon refining mesh.

\subsection{Effect of irradiation and strain level}

\begin{figure}[H]
\centering
\subfigure[]{\includegraphics[height = 5.5cm]{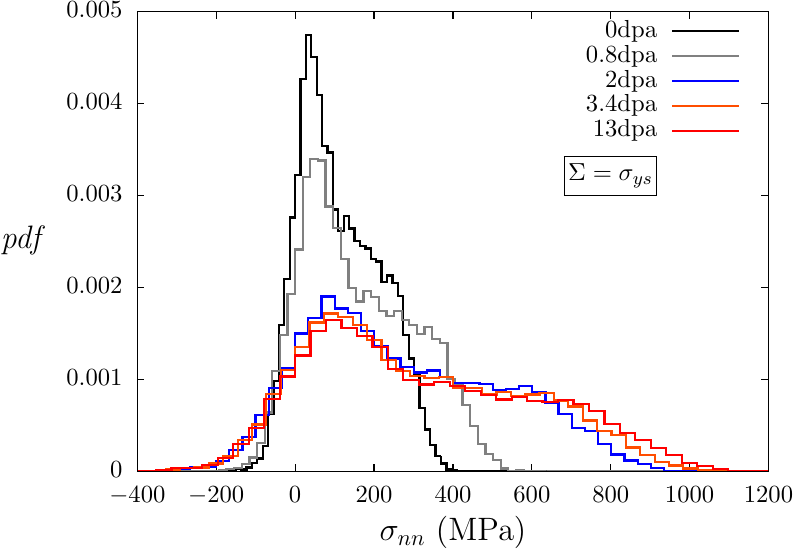}}
\subfigure[]{\includegraphics[height = 5.5cm]{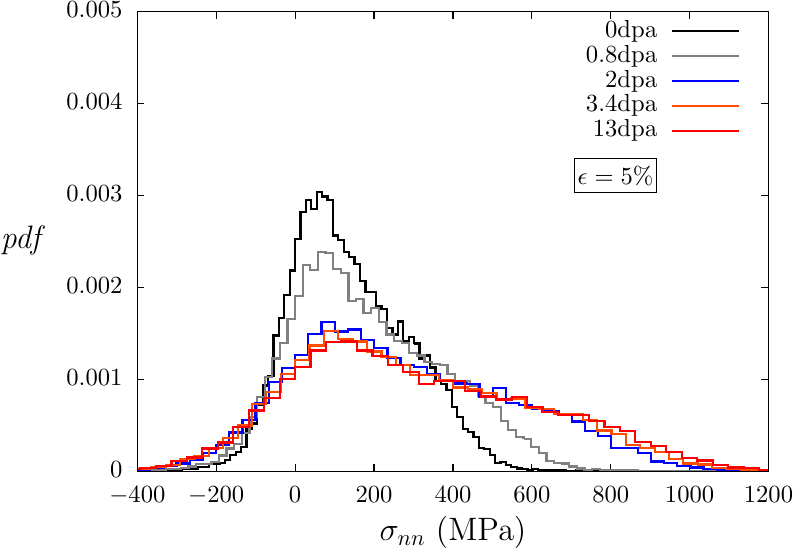}}
\caption{
Probability density functions of intergranular normal stress
$\sigma_{nn}$ as a function of the level of irradiation, for (a)
macroscopic yield stress and (b) a total strain of $5\%$.}
\label{fig:gra1}
\end{figure}

Probability density functions of normal intergranular stress are shown
on Fig.~\ref{fig:gra1} as a function of irradiation level, at yield
stress (Fig.~\ref{fig:gra1}a) and for a strain of $5\%$
(Fig.~\ref{fig:gra1}b). The distributions have complex shapes, resulting from elastic and/or plastic incompatibilities between adjacent grains. One may note here that such shapes are not Gaussian, which means that using only standard deviation as done in other studies to describe them is not sufficient. The higher the macroscopic stress (yield stress increases with irradiation level, see Fig. 1c),
the wider the pdf of normal intergranular stress, and the larger the upper tail of the distribution as can be seen on Fig.~\ref{fig:gra1}a. As the level of
irradiation increases, the $\textit{pdf}$ flattens, and a significant
probability (defined as the integral of the $\textit{pdf}$) is always associated with stresses higher than the macroscopic value of the stress. The difference between unirradiated and
irradiated material is reduced as the strain level increases, which
may be related to the fact that the unirradiated material has more
work-hardening capabilities than irradiated material, which tends to bring stress states closer for a given strain value.

\begin{figure}[H]
\centering
\subfigure[]{\includegraphics[height = 5.5cm]{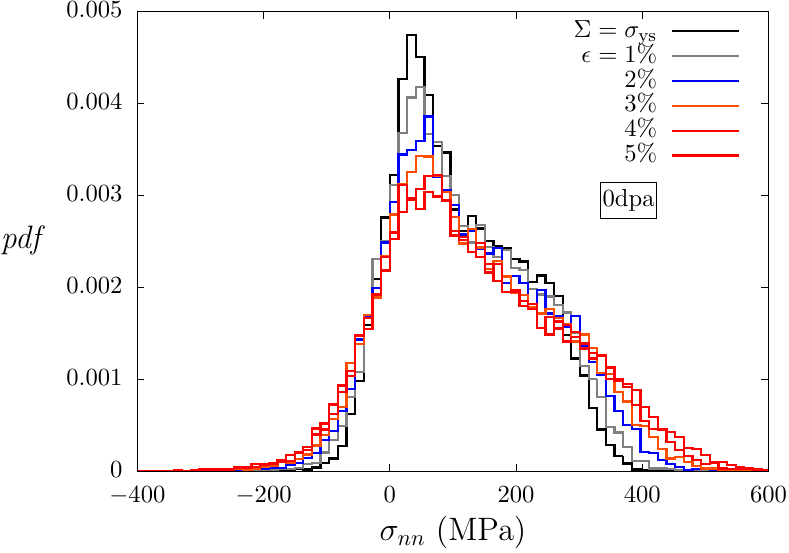}}
\subfigure[]{\includegraphics[height = 5.5cm]{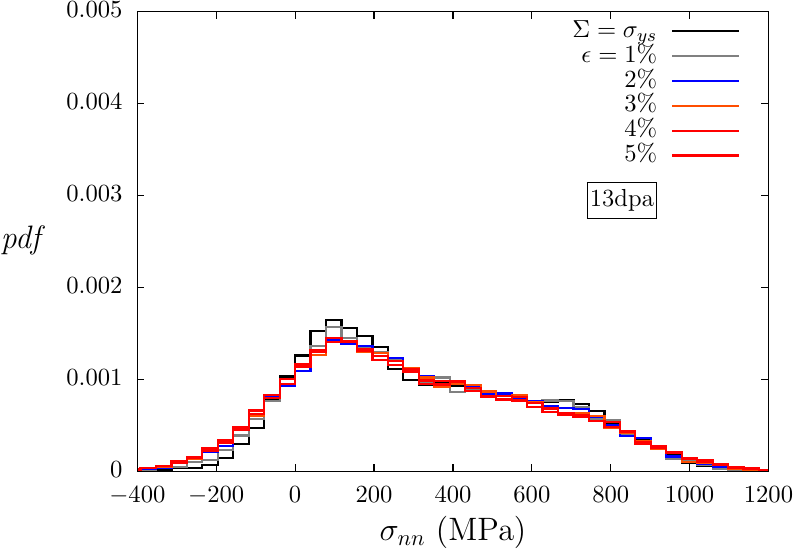}}
\caption{
Probability density functions of intergranular normal stress
$\sigma_{nn}$ as a function of strain, for (a) unirradiated material
(b) 13 dpa irradiated material.}
\label{fig:gra2}
\end{figure}

In addition, for two different levels of irradiation, the evolution of the
intergranular normal stress \textit{pdf} with total strain is shown on
Fig.~\ref{fig:gra2}. For the unirradiated material, the \textit{pdf}
flattens with work-hardening, while there are 
no changes for
highly irradiated material that has almost no work-hardening capabilities, and thus no increase of the upper tail of distribution compared to unirradiated material.

\subsection{Master curve}

\begin{figure}[H]
\centering
\includegraphics[height = 5.5cm]{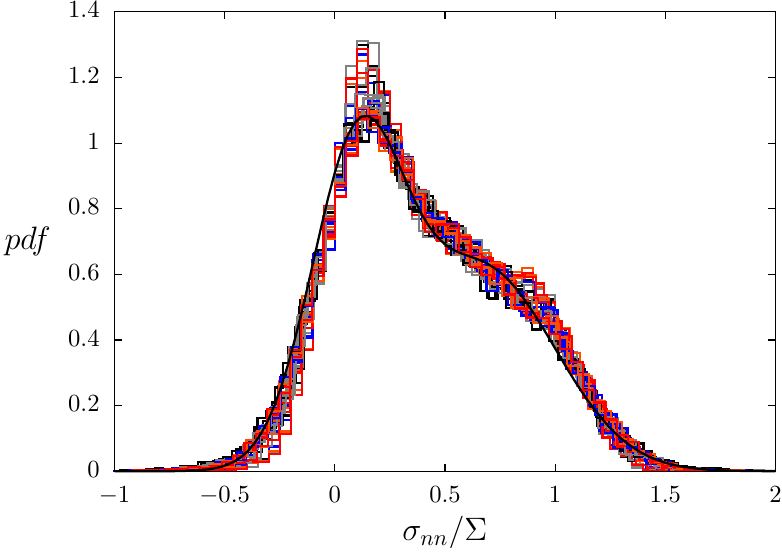}
\caption{
Probability density functions of intergranular normal stress
$\sigma_{nn}$ rescaled by macroscopic stress $\Sigma$. All curves,
spanning 0 to 5\% strain, and 0 to 13 dpa, are well approximated by a
mixture of two normal distributions (Eq.~\ref{eqmc}) with parameters
$a=0.51$, $\mu_1 = 0.67$, $\sigma_1 = 0.33$, $\mu_2 = 0.10$ and
$\sigma_2 = 0.21$ (black thick line).}
\label{fig:gra3}
\end{figure}

Rescaling all \textit{pdf}s by the macroscopic uniaxial stress $\Sigma$ at
which they were obtained\footnote{For the relatively small values of global strain considered in this study, we do not differentiate engineering stress and true stress.}, for all strains (yield strain-5\%) and
irradiation levels (0-13 dpa), leads at first order to a master curve
shown on Fig.~\ref{fig:gra3}. This master curve appears to be well
approximated by a mixture of two normal distributions

\begin{equation}
\mathrm{pdf\left(x = \frac{\sigma_{nn}}{\Sigma}   \right)} = \frac{a}{\sqrt{2\pi}\sigma_1}
\exp{\left[-\frac{(x-\mu_1)^2}{2\sigma_1^2}\right]} +
\frac{1-a}{\sqrt{2\pi}\sigma_2}
\exp{\left[-\frac{(x-\mu_2)^2}{2\sigma_2^2}\right]}
\label{eqmc}
\end{equation}

\noindent
with parameters
$a=0.51$, $\mu_1 = 0.67$, $\sigma_1 = 0.33$, $\mu_2 = 0.10$ and
$\sigma_2 = 0.21$.  Such kind of distribution of normal stress is found for example for the sum of 
an uniaxial stress tensor $\Sigma\ \bm{\underline{e}_3}
\otimes\bm{\underline{e}_3}$ and a fluctuation diagonal stress tensor,
$\bm{\underline{\sigma}} = \sigma_{imp}^1\ \bm{\underline{e}_1}
\otimes\bm{\underline{e}_1} + \sigma_{imp}^2\ \bm{\underline{e}_2}
\otimes\bm{\underline{e}_2} + \left(\Sigma + \sigma_{imp}^3\right)\ \bm{\underline{e}_3}
\otimes\bm{\underline{e}_3}$,
assuming random normal orientations of the grain boundary facets,
where $\sigma_{imp}^i$ follows a normal distribution with zero
mean. Such stress tensor can be seen as an approximation of stress at
grain boundaries inside a polycrystalline aggregate in uniaxial
tension, where $\sigma_{imp}^i$ comes from incompatibilities of
deformations between adjacent grains, which explains the shapes of
distributions obtained throughout this
study.
Additional simulations have shown in fact that the shape of this master curve, and in particular the upper tail, remains practically unchanged when plasticity is omitted so that only elastic interactions are assumed between the grains, as simulations using only cubic elasticity (and no plasticity) give close results. However, for larger strains, significant deviations from cubic elasticity results are expected, but may not be relevant regarding IGSCC of post-irradiated materials where macroscopic stress is close or below the yield stress. For example, at constant load tests  macroscopic stress is equal or below yield stress \cite{freyer} and at constant extension rate tests strain is usually lower than few percents \cite{lemillier}.

In order that the probability density functions obtained in this study
to be useful for IGSCC modelling, the dependence to the refinement of
the mesh and grain number need to be assessed, which is done in the
next section. Sensitivity to crystallographic texture, Grain shapes
and to free surface is also studied to uncover the relevant parameters
governing quantitatively the master curve, and especially the upper
tail of the distributions which is relevant for IGSCC modelling from a
statistical point of view.

\section{Master curve parameters assessment}
\label{sensitivitystudy}

The results shown in Section~\ref{interdis} (referred to as
\textit{coarse wire}) are compared to the ones obtained with other
aggregates presented in Table~2. Fine and coarse wires differ
only by the refinement of the mesh and are based on wire
microstructure (both for the grain shapes and orientations, thus with
crystallographic texture), while VORO-216 and VORO-343 are Voronoi
tesselation based aggregates with different number of grains, with no
crystallographic texture.

\subsection{Grain number and mesh density}

\begin{figure}[H]
\centering
\subfigure[]{\includegraphics[height = 5.5cm]{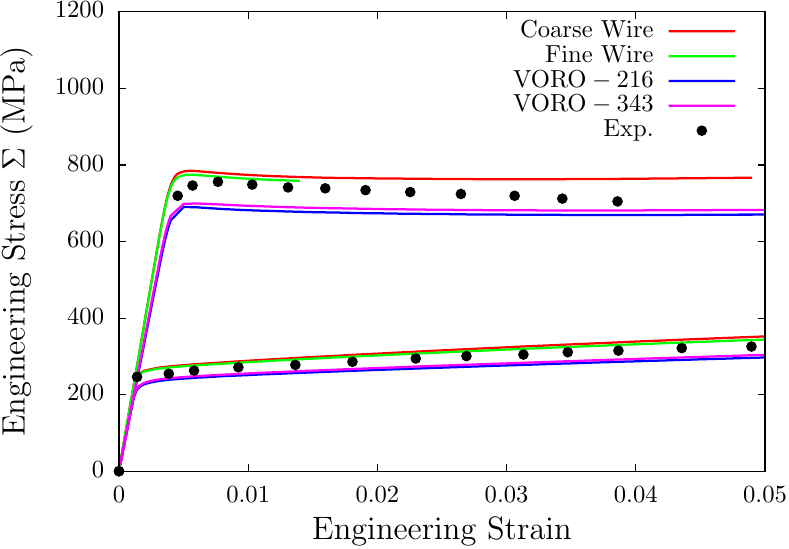}}
\hspace{1cm}
\subfigure[]{\includegraphics[height = 4.5cm]{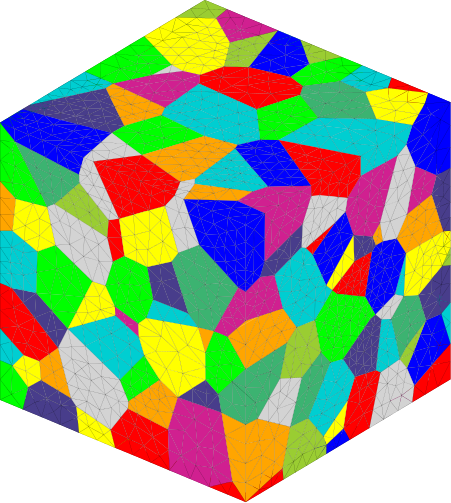}}
\caption{
(a) Comparisons of macroscopic strain-stress curves between experimental
data and numerical simulations for 0dpa and 13dpa. (b) Typical Voronoi aggregate mesh used in this study.}
\label{fig:gra7}
\end{figure}

The macroscopic stress-strain curves for the four aggregates
are shown on
Fig.~\ref{fig:gra7} for two levels of irradiation (0, 13dpa), and compared to
experimental data on which the constitutive equations were calibrated
\cite{xuhanconf}.
Macroscopic convergence is achieved for the wire aggregates, and the
results for Voronoi aggregates indicate that grain number convergence
is achieved for 216 grains. All numerical curves are in reasonable
agreement with the experimental data. Voronoi aggregates appear to be
softer, which can be explained by the fact that the parameters of the
constitutive law were adjusted in \cite{xuhanconf} on Voronoi models
where the nodes on the lateral surfaces of the aggregate were imposed
to move by the same amount. In this way, the interaction with the
surrounding material was simulated to provide a bulk response. This
kind of boundary conditions leads to harder mechanical
response. Voronoi and Wire aggregates used here have same boundary
conditions, the difference between the two arises from the difference in crystallographic texture as mentionned previously.

\begin{figure}[H]
\centering
\subfigure[]{\includegraphics[height = 5.5cm]{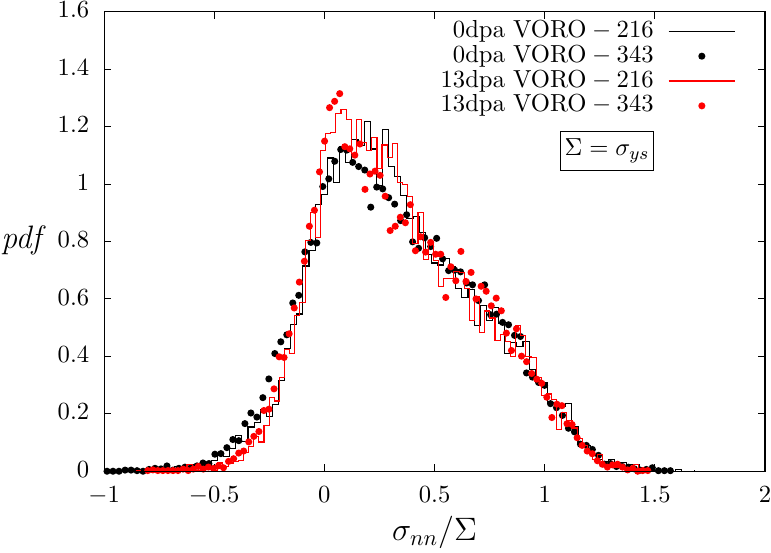}}
\subfigure[]{\includegraphics[height = 5.5cm]{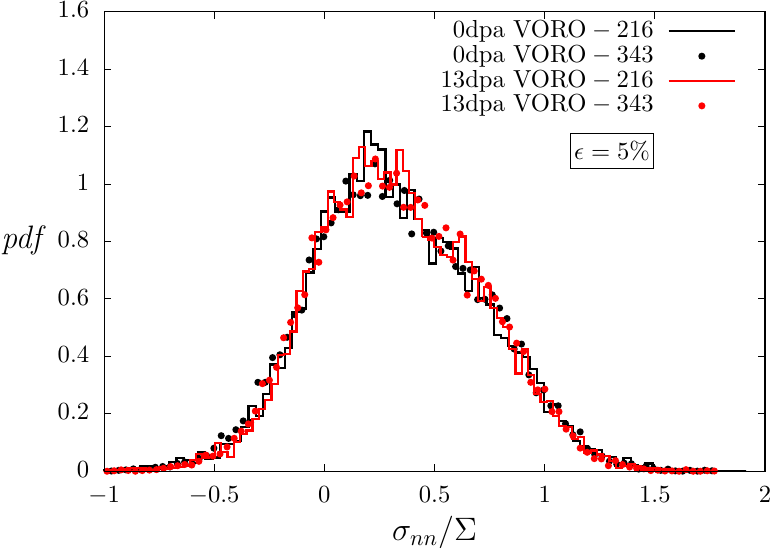}}
\caption{
Probability density functions of intergranular normal stress
$\sigma_{nn}$ rescaled by the macroscopic stress for Voronoi
aggregates, for (a) yield stress (b) strain of $5\%$. Simulations with
coarser mesh (not presented here) have confirmed the mesh
convergence.}
\label{fig:gra8}
\end{figure}

\begin{figure}[H]
\centering
\subfigure[]{\includegraphics[height = 5.5cm]{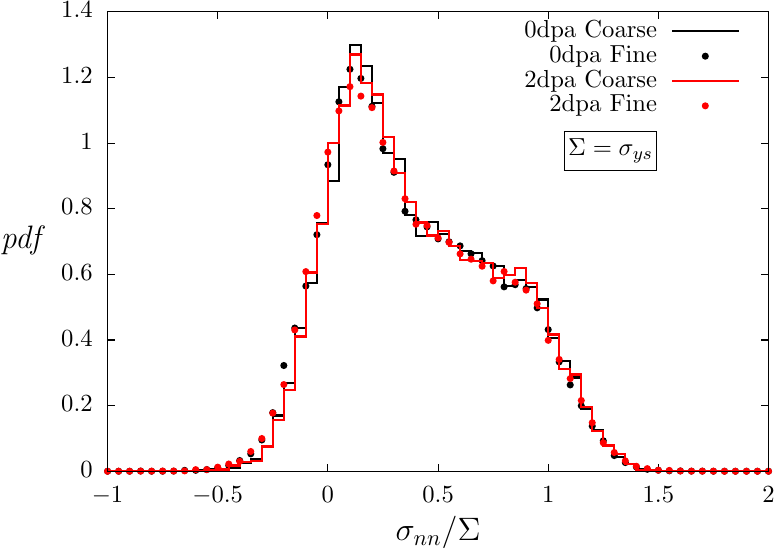}}
\subfigure[]{\includegraphics[height = 5.5cm]{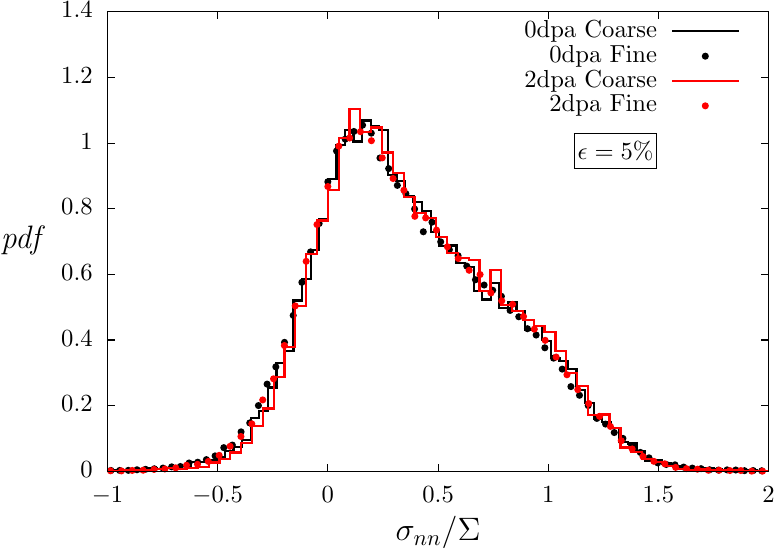}}
\caption{
Probability density functions of intergranular normal stress
$\sigma_{nn}$ rescaled by the macroscopic stress for wire
aggregates, for (a) yield stress (b) strain of $5\%$.}
\label{fig:gra5}
\end{figure}

The local convergence with respect to the number of grains is shown on
Fig.~\ref{fig:gra8}, where the probability density functions of
intergranular normal stress are shown for different strain and
irradiation levels. Simulations on the finer mesh for the wire
aggregate (Fig.~\ref{fig:gra5}) show mesh convergence and also
validate the methodology to compute stresses at grain boundaries.

\subsection{Texture and Grain shape effects}

\begin{figure}[H]
\centering
\subfigure[]{\includegraphics[height = 5.5cm]{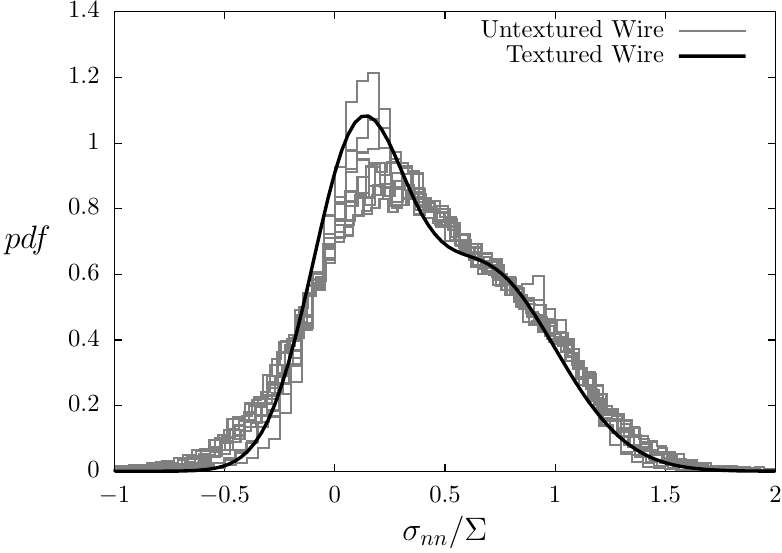}}
\subfigure[]{\includegraphics[height = 5.5cm]{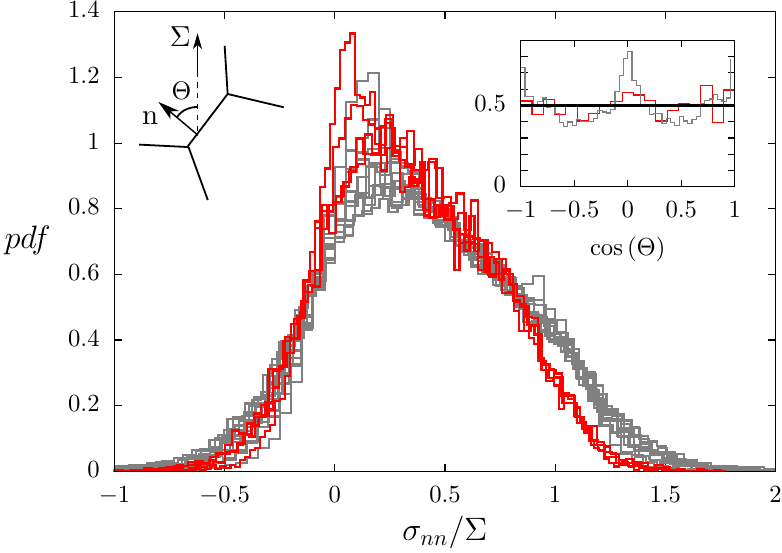}}
\caption{
Comparisons of probability density functions of intergranular normal
stress rescaled by macroscopic stress (a) obtained on wire model with random crystallographic orientations (solid gray lines) to the master curve (black line) obtained for original wire aggregate (b) of wire model with random crystallographic orientations (solid gray lines) to voronoi model (red lines) for different levels of strain and irradiation. Inset: Probability density function of
the cosine of the angle between the grain boundary normal and the
loading direction, for a typical Voronoi aggregate (red line), coarse
wire aggregate (gray line) compared to the theoretical prediction
assuming random orientations (black line).}
\label{fig:gra3b}
\end{figure}

As previously mentioned, the wire exhibits a non-zero texture. Two regions of predominant crystallographic orientations are visible (Fig.~2), one close to [111] and the
other close to [001] orientation. Such a finite (but small) texture is
a consequence of the manufacturing process of the wire. To assess the
influence of the texture, additional simulations were performed on the
wire aggregate with random crystallographic orientations.
As shown on Fig.~\ref{fig:gra3b}a, significant deviations are observed, mainly close to zero stress, and thus full distributions depends significantly on crystallographic texture. However, the upper tail is left unchanged and appears to be relatively insensitive to small deviations from random orientations. In addition, results obtained on wire aggregate with random crystallographic orientations are compared to those obtained on Voronoi aggregates (Fig.~\ref{fig:gra3b}b).
Significant deviations are observed for the upper tail, wire aggregate showing larger intergranular stress than Voronoi aggregate.
These are analysed further in the inset of Fig.~\ref{fig:gra3b}b. A
clear deviation from the uniform probability density distribution of
the cosine of the angle, $\cos(\Theta)$, between the grain boundary
normal and the loading direction observed in the wire suggests that
grain boundary facets that are parallel,
$\cos(\Theta)\sim \pm 1$, or perpendicular, $\cos(\Theta)\sim 0$, to
the two base surfaces of the wire are relatively more frequent. This kind of distribution of grain
shapes in a realistic wire model also seems to be a consequence of the
manufacturing process. As grain boundaries perpendicular to loading direction are more susceptible to generate large stress, the difference between Voronoi and wire aggregate is attributed to differences in grain shapes.

\subsection{Free surface effect}

\begin{figure}[H]
\centering
\includegraphics[height = 5.5cm]{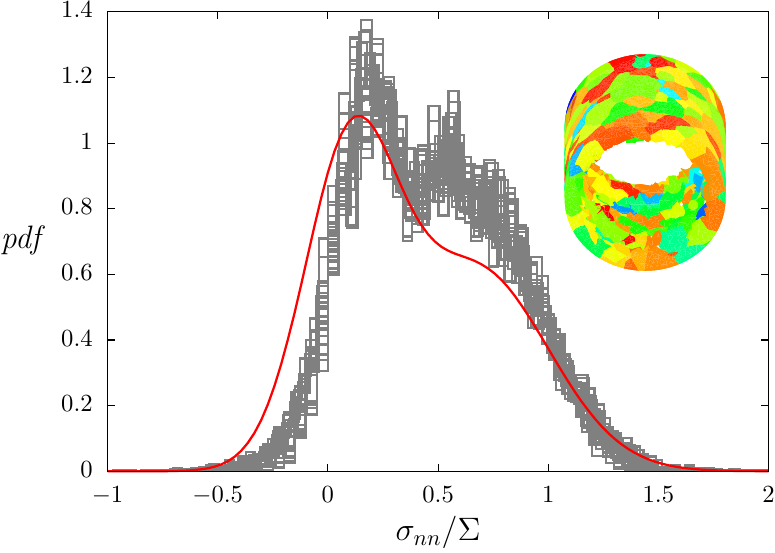}
\caption{
Comparison of probability density functions of intergranular normal
stress rescaled by macroscopic stress for wire aggregate (solid red line, see Fig.~\protect\ref{fig:gra3}, considering the entire
aggregate) and restricting to grain boundaries close to the free
surface ($r/R > 0.90$, where $r$ is the radial coordinate and $R$ the radius of the wire. Lines for various strain and irradiation levels are shown in gray.}
\label{fig:gra4}
\end{figure}

All probability density functions presented here above were obtained
after post-processing entire aggregates in uniaxial tension,
\textit{i.e.} independently of the location of the boundary between
the two grains with respect to the distance from the free surface. As
cracking coming from IGSCC phenomenon is a damage process initiating at the surface, the \textit{pdf} of intergranular normal stress
close to the free surface have also been studied. Comparison 
between the master curve of Fig.~\ref{fig:gra3} to \textit{pdf}
obtained on the wire aggregate by considering only grain
boundaries close to the free surface is presented on Fig.~\ref{fig:gra4}.

Probability density functions are significantly altered when looking
at the bulk or close to the free surface. However, the upper
tail is not significantly changed. As cracking initiation prediction will mainly
depend on the higher level of stress occurring on the sample, this
figure indicates that results obtained in bulk aggregates provides a good description of stress distribution close to a free surface, and therefore be used to
model IGSCC.

\subsection{Intergranular normal stress \textit{vs.} plastic incompatibilities}

\begin{figure}[H]
\centering
\includegraphics[height = 7.5cm]{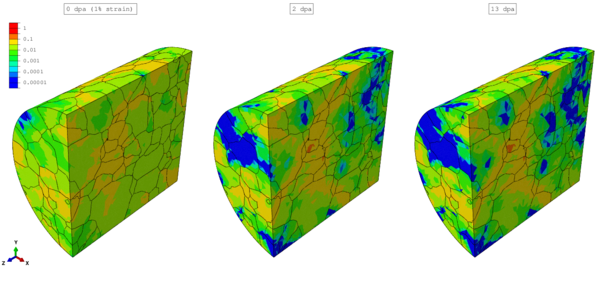}
\caption{
Comparison of the amount of plastic slip (as measured by $\sum_\al
|\gamma^{\alpha}|$) for 1\% strain for different irradiation
levels. Half of the wire is omitted to visualize better the interior
slip distribution. Logarithmic scale is used in the legend.}
\label{fig:gamsum}
\end{figure}

As discussed in section 3.2, the stress distributions obtained throughout this study up to 5\% strain were found to be rather insensitive to the presence of plasticity, suggesting that in stainless steel both elastic and plastic anisotropies develop similar incompatibilites between the grains. In order to assess this observation more precisely, relation between normal intergranular stress and plastic slip at grain boundary is computed. First, cumulative total plastic slip is presented on Fig.~\ref{fig:gamsum} at
different irradiation levels for a given strain, showing large
variations due to orientation of the grains with respect to the
loading direction. At a given total strain, the homogeneity of  plastic slip activity is reduced with irradiation, increasing material volumes that do not strain at all (blue regions in Fig.~11).

\begin{figure}[H]
\centering
\subfigure[]{\includegraphics[height = 5.5cm]{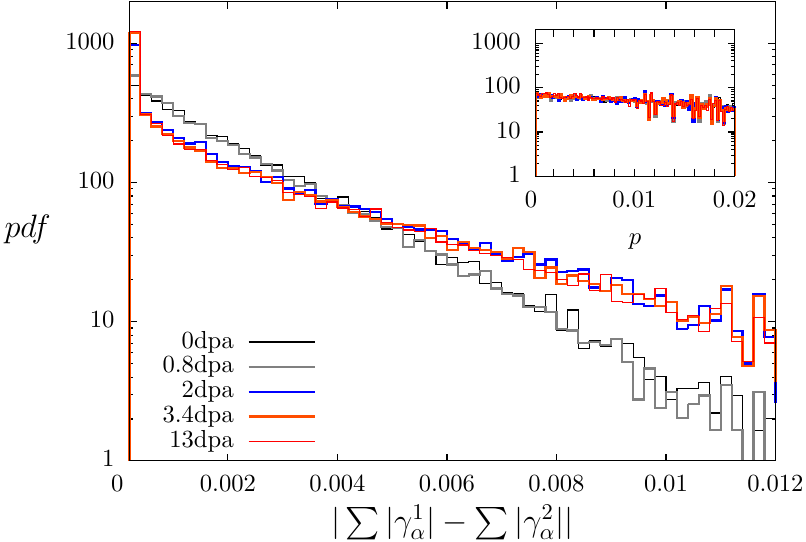}}
\subfigure[]{\includegraphics[height = 5.5cm]{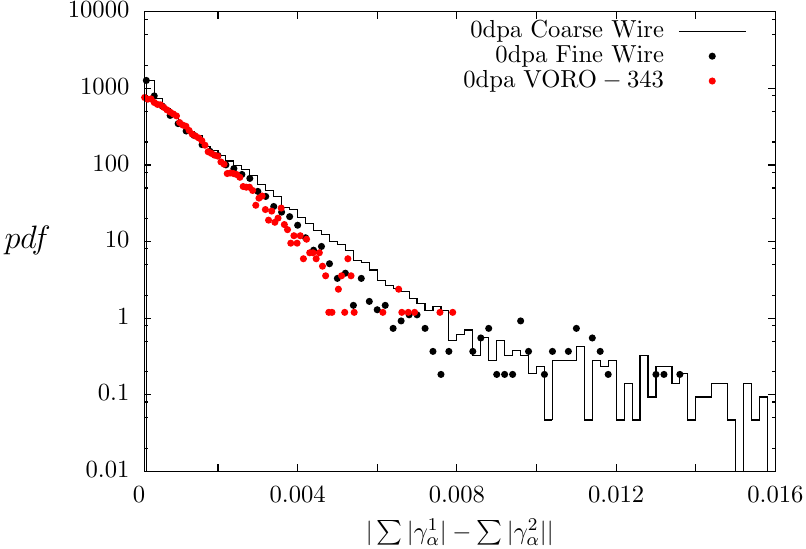}}
\caption{
(a) Probability density functions of difference of cumulative plastic slips at grain boundaries at macroscopic yield stress (Inset: for 5\% total strain), for different irradiation levels for wire aggregate (b) Probability density functions of difference of cumulative plastic slips at grain boundaries at macroscopic yield stress for unirradiated material. Comparison between wire and Voronoi aggregates.}
\label{fig:gra1011}
\end{figure}

Plastic strain incompatibilities between adjacent grains are measured through the difference of cumulative plastic slips (defined as $\sum_{\alpha} |\gamma_{\alpha}|$) at grain boundaries\footnote{On each side of the grain boundary, cumulative plastic slip is computed on the closest Gauss points.} and shown on Fig.~\ref{fig:gra1011}. While some differences may be observed at yield stress for the different irradiation levels (Fig.~\ref{fig:gra1011}a), distributions tend to become similar at higher strains (inset Fig.~\ref{fig:gra1011}a). These distributions were observed to be rather insensitive to mesh and details of the microstructure.\\

\begin{figure}[H]
\centering
\includegraphics[height = 9cm]{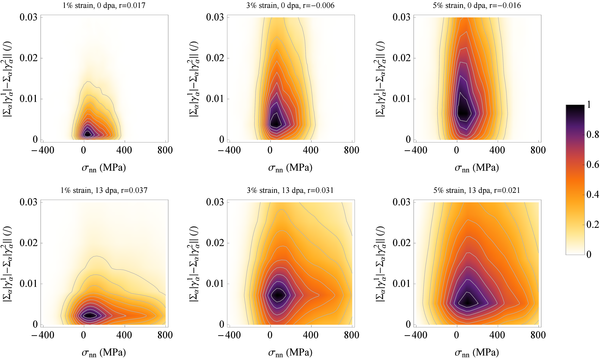}
\caption{2D probability density functions of intergranular stress \textit{vs.} plastic slips discontinuities, for unirradiated and irradiated (13dpa) material, for three different levels of strain (1\%, 3\% and 5\%). Small linear correlation coefficients, r, indicate a vanishing correlation between the two quantities.}
\label{fig:2Dhist}
\end{figure}

Finally, 2D probability density functions of intergranular normal stress \textit{vs.} of plastic slips discontinuities are shown on Fig.~\ref{fig:2Dhist}. With the constitutive equations and for the strain level used in this study, there is no correlation between high intergranular normal stress and high plastic slips discontinuities.

\section{Discussion}
\setlength{\itemsep}{0pt} 
Predicting IGSCC of post-irradiated stainless steels requires evaluation of local stresses at grain boundaries. Finite-element simulations have been performed in this study on a realistic aggregate and using crystal plasticity constitutive equations calibrated for irradiated stainless steel to compute probability density functions of intergranular normal stresses at grain boundaries.
As intergranular normal stresses used to compute probability density
functions were weighted by the surface of the boundary on which they
were obtained, the probability to have stress $\sigma_{nn}$ higher
than a given threshold $\sigma_c$: $P(\sigma_{nn} \geq \sigma_c) =
\int_{\sigma_c}^{+\infty} pdf(\sigma_{nn})d\sigma_{nn}$ also
corresponds to the percentage of surface of grain boundaries subjected
to a stress higher than $\sigma_c$, and also approximately to the
percentage of grain boundary length at the free surface subjected to a
stress higher than $\sigma_c$, as shown on Fig.~\ref{fig:gra4} for
sufficiently high $\sigma_c$. Application of these distributions to the modelling of IGSCC requires the determination of the strength of the grain boundary as a function of oxidation level and irradiation in LWR environment. Therefore, once critical stress for
oxidised grain boundary in LWR environment would become available in
the literature, uncoupled modelling of IGSCC based on the
distributions obtained in this study would become available, by
setting a critical fraction of fractured grain boundaries above which
global fracture of the specimen considered is expected. Such uncoupled
modelling has been for example used successfully in ductile fracture analysis based
on critical void radius and Rice-Tracey law \cite{besson2010}.

Numerous studies have been conducted in the recent past on IGSCC of (ion-irradiated) austenitic stainless steel, mainly with applied stress well above yield stress. It has been shown (see, \textit{e.g.}, \cite{mdmcm}) that dislocation channelling (or clear bands) resulting in strain localization at the grain scale may be a factor controlling IGSCC of irradiated stainless steel. The constitutive equations used in this study do not display instabilities and subsequent strong strain localization at the grain scale, and therefore the results and methodology proposed in this study would probably show limitations when applied to highly strained materials. However, and despite the fact the stress at which clear bands appear is still a matter of debate, the results shown here are expected to be suitable for low macroscopic stress (up to yield stress) and/or low irradiation level, for which channelling dislocation mode is not prominent, or for highly sensitized grain boundary. This is supported by results shown in Fig.~\ref{fig:2Dhist}, showing no correlation between high intergranular stress and plastic slip incompatibilities, indicating that a stress-controlled failure mode could appear before large plastic incompatibilities, for sufficiently weak grain boundary.

\section{Conclusion}
\setlength{\itemsep}{0pt}  

IGSCC modelling of irradiated stainless steel requires evaluation of
stress distributions at grain boundaries and their evolution with both
strain and irradiation level. Normal stress
distributions at grain boundaries for (un)-irradiated austenitic
stainless steel have been provided based on finite elements simulations performed on
realistic aggregates with recently proposed physically-based crystal plasticity constitutive equations. In the realistic aggregate a mesh is obtained
from diffraction contrast tomography data of an austenitic stainless
steel wire, that gives both grain shapes and crystallographic
orientations. Crystal plasticity constitutive equations are based on
physically-based modelling, and have been adjusted so as to reproduce
both macroscopic tensile tests and microscopic evolution of
microstructure variables (dislocations and defects densities).

Once rescaled by the macroscopic stress, the distributions obtained
are found to be well approximated by a master curve in the range [0-13
dpa] and [yield-5\% strain]. The upper tail of this master curve, relevant for IGSCC initiation modelling, does not depend
strongly on the crystallographic texture of the material close to
random orientations, but is sensitive to grain shapes. This implies
that realistic aggregates are required to obtain relevant intergranular stress distributions. In addition, free surface effect does not affect significantly the distributions of upper tail, thus allowing to use the distributions obtained in this study to describe intergranular stress at the free surface, which is relevant for IGSCC initiation
. Based on these results, a methodology for uncoupled modelling of IGSCC of irradiated materials has been proposed and anticipated to provide reliable estimations for low strain and sufficently weak grain boundaries, which are expected to be encountered in highly irradiated stainless steels.\\

\noindent{\bf Acknowledgments\\}\\
The authors acknowledge the financial support from Slovenian Research
Agency and French Atomic Energy Commission through the bilateral
project “Comprehensive and reliable prediction of LWRs internals
mechanical behaviour based on microstructure-informed modeling” between
CEA and JSI in years 2013-2014. J.H would like to thank Chao Ling,
Thomas Helfer and Lionel G\'el\'ebart for fruitful discussions and
technical assistance in finite element simulations.

\newpage
\section{Appendix}
\subsection{Abaqus implementation}

The implementation of constitutive equations in ABAQUS is done through a user-material
subroutine UMAT \cite{abaqus}. Following Huang's implementation
\cite{huang}, large deformation theory is used along with forward
gradient time integration scheme ($\theta$ method) and linear
incremental formulation. The implementation results in a semi-implicit
scheme. Detailed formulation of the implementation can be found in
\cite{nene2014}.

A time integration scheme assumes a linear relation among the
increments of stresses, strains and internal variables such as shear
strains, dislocation densities and irradiation defect (Frank loop)
densities. The stresses and internal variables, $v\in\left\{\gm^\al,
r_D^\al, r_L^p\right\}$, are evaluated at the start of the increment
so that
\be
  v_{t+\Dl t}=v_t + \Dl v_{t}
\ee
where the increment of the internal variable $v$ within the time
increment $\Dl t$ is defined as
\be
  \Dl v_t = v(t+\Dl t) - v(t) = \Dl t \left((1-\theta)\dot{v}_t + 
  \theta\dot{v}_{t+\Dl t}\right).
  \label{eq_00}
\ee
A parameter $\theta$, ranging from 0 to 1, is introduced to employ a
linear interpolation within $\Dl t$. A value of $\theta=0$ corresponds
to explicit Euler time integration scheme and $\theta>0$ to
semi-implicit one. A choice of $\theta$ between 0.5 and 1 is
recommended \cite{huang}.

A time derivative at $t+\Dl t$ is further approximated by applying the
first-order Taylor expansion
\be
  \dot{v}_{t+\Dl t} = \dot{v}_{t} + \sum_i \frac{\partial\dot{v}}{\partial x_i}\Dl x_i
  \label{eq_dgm0}
\ee
where partial derivatives are taken with respect to all time-dependent variables $x_i$ that influence $\nu$, and $\Delta x_i$ is the corresponding increment within $\Delta t$. In Eq. (1), for example, two variables apply, $x_1=\tau^\alpha$ and $x_2=\tau_c^\alpha$. Rearranging the above equations
Eqs. (\ref{eq_00} -- \ref{eq_dgm0}) gives the following incremental
relation
\be
  \Dl v_t = \Dl t \left(\dot{v}_t + \theta\sum_i\frac{\partial\dot{v}}{\partial x_i}\Dl x_i\right).
  \label{eq_dgm}
\ee
All partial derivatives in incremental relations for $\Dl v_t$ can be
explicitly derived from constitutive relations, Eqs. (\ref{eq_gm} --
\ref{eq_rl}), and evaluated at the start of the increment. Also, each
$\Dl x_i$ may be expressed as a linear function of the increments
$\Dl\gm^\al, \Dl r_D^\al, \Dl r_L^p$ and strain increment
$\Dl\varepsilon_{ij}$ \cite{nene2014}. In this way, a linear system of
equations can be set up for the 28 unknowns $\Dl\gm^\al, \Dl r_D^\al,
\Dl r_L^p$. The solution for the increments is obtained by finding the
inverse of the $28\times 28$ matrix using the standard LU
decomposition with Gauss elimination.

\subsection{Cast3M implementation}

Numerical simulations performed in Cast3M use a fully implicit
implementation of the constitutive equations generated by the MFront
code generator \cite{mfront}. 
The implicit scheme assumes that stresses, strains and all internal
variables, $v\in\left\{E^{ij}, \gm^\al, r_D^\al, r_L^p\right\}$, are
evaluated at the end of time increment, $t+\Dl t$, so that
\be
  v_{t+\Dl t}=v_t + \Dl v_{t+\Dl t}.
\ee
Note that elastic Green-Lagrange strain tensor $E$ is considered here
as additional free internal variable to avoid the risk of inaccuracy
in the numerical calculation of the elastic and plastic part of
deformation gradient, $F^e$ and $F^p$ \cite{phdHan}. The elastic part
of the deformation at the end of the time step is approximated by the
formula

\ba
F^e_{t+\Dl t} = \Delta F \  F^e_{t} \left( \Delta F^p \right)^{-1} \ \ \ \ \ \ \ \mathrm{with} \ \ \ \ \ \ \ \left( \Delta F^p \right)^{-1} \approx \frac{1 - \sum_{\alpha=1}^{12} \Delta \gamma^{\alpha} N^{\alpha}}{\left[\det{\left(1 - \sum_{\alpha=1}^{12} \Delta \gamma^{\alpha} N^{\alpha}\right)}    \right]^{1/3}}
\label{eq_fp}
\ea
using a multiplicative decomposition of the deformation gradient,
$F=F^e F^p$, and $\Delta F = F_{t+\Dl t} F_{t}^{-1}$.
To integrate internal variables at $t+\Dl t$, a nonlinear system of
equations is solved by a Newton-Raphson iteration method. In this
respect, vanishing residual functions, $R_{E^{ij}}, R_{\gm^\al},
R_{r_D^\al}, R_{r_L^p}$, are defined by the
evolution relations, Eqs. (\ref{eq_gm} -- \ref{eq_rl}), and from
Eq. (\ref{eq_fp}),
\ba
  &&R_{E^{ij}}=\Dl E^{ij} +E^{ij}_t - \frac{1}{2}\left(^tF^e_{t+\Dl t}F^e_{t+\Dl t}   -1 \right)\nonumber\\
  &&R_{\gm^\al}=\Dl\gm^\al - \left\langle\frac{|\tau^\al| - 
                  \tau_c^\al}{K_0}\right\rangle^n {\rm sign}(\tau^\al)\Dl t\nonumber\\
  &&R_{r_D^\al} = \Dl r_D^\al - \left\{ \begin{array}{ll}
  \left(\frac{1}{\kp}\sqrt{\sum\limits_{\bt=1}^{12} b^{\al\bt}r_D^\bt}-G_c r_D^\al
  \right)|\Dl\gm^\al| &; \hbox{non-irradiated}\nonumber\\
  \left(\frac{1}{\kp}\sqrt{\sum\limits_{\bt=1}^{12} b^{\al\bt}r_D^\bt}+
  \frac{1}{\kp}\sqrt{K_{dl}\sum\limits_{p=1}^4 r_L^p}-G_c r_D^\al\right)|\Dl\gm^\al| &; \hbox{irradiated}
  \end{array}\right. \nonumber\\
  &&R_{r_L^p} = \Dl r_L^p + A_L (r_L^p-r_L^{sat})\left(\sum_{\al\in {\rm plane}\ p}^3\!\!\!\! r_D^\al\right) 
  \left(\sum_{\al\in {\rm plane}\ p}^3\!\!\!\! |\Dl \gm^\al|\right).
\ea


\bibliographystyle{elsarticle-num.bst}
\bibliography{spebib2}

\end{document}